%% file: elsarticle-template.tex
\documentclass[preprint,10pt]{elsarticle}
\usepackage[inner=3cm,outer=2cm]{geometry}
\usepackage{lineno,hyperref}

\usepackage{mathtools}
\usepackage{amsmath}
\usepackage{float}
\usepackage{amssymb}
\usepackage{multirow}
\usepackage{tabularx}
\usepackage{caption}
\usepackage{subcaption}
\usepackage{graphicx}
\usepackage{setspace}
\usepackage{titlesec}

\newcommand{\probability}[2]{%
  \left\{\begin{array}{@{}l<{\kern-\nulldelimiterspace}@{}}
    #1
    \end{array}\right. \qquad  #2
} 
\usepackage{tikz}
\usepackage[framemethod=default]{mdframed}

\titlespacing{\section}{0pt}{*2}{*1}
\titlespacing{\subsection}{0pt}{*1.5}{*1}
\titlespacing{\subsubsection}{0pt}{*1.5}{*0.8}

\journal{Journal of Network and Computer Applications}

\begin{document}

\begin{frontmatter}
\title{Secure Multi-Cloud Virtual Network Embedding}

\author[lasigeAddress]{Max Alaluna\corref{mycorrespondingauthor}}
\cortext[mycorrespondingauthor]{Corresponding author}
\ead{malaluna@lasige.di.fc.ul.pt}
\author[lasigeAddress]{Lu\'is~Ferrolho}
\ead{lferrolho@lasige.di.fc.ul.pt}
\author[istAddress]{Jos\'e~Rui~Figueira}
\ead{figueira@tecnico.ulisboa.pt}
\author[lasigeAddress]{Nuno~Neves}
\ead{nuno@di.fc.ul.pt}
\author[lasigeAddress]{Fernando~M.~V.~Ramos}
\ead{fvramos@ciencias.ulisboa.pt}

\address[lasigeAddress]{LASIGE, Faculdade de Ci\^{e}ncias, Universidade de Lisboa, Departamento de inform\'{a}tica, Edif\'{i}cio C6 Piso 3, Campo Grande, 1749-016 Lisboa, Portugal}
\address[istAddress]{CEG--IST, Instituto Superior T\'ecnico, Universidade de Lisboa, Portugal}









\begin{abstract}
Modern network virtualization platforms enable users to specify custom topologies and arbitrary addressing schemes for their virtual networks.
These platforms have, however, been targeting the data center of a single provider, which is insufficient to support (critical) applications that need to be deployed across multiple trust domains, while enforcing diverse security requirements.
This paper addresses this limitation by presenting a novel solution for the central resource allocation problem of network virtualization -- the virtual network embedding, which aims to find efficient mappings of virtual network requests onto the substrate network.
We improve over the state-of-the-art by considering security as a first-class citizen of virtual networks, while enhancing the substrate infrastructure with resources from multiple cloud providers.
Our solution enables the definition of flexible policies in three core elements: on the virtual links, where alternative security compromises can be explored (\textit{e.g.},  encryption); on the virtual switches, supporting various degrees of protection and redundancy if necessary; and on the substrate infrastructure, extending it across multiple clouds, including public and private facilities, with their inherently diverse trust levels associated.
We propose an optimal solution to this problem formulated as a Mixed Integer Linear Program (MILP).
The results of our evaluation give insight into the trade-offs associated with the inclusion of security demands into network virtualization.
In particular, they provide evidence that enhancing the user's virtual networks with security does not preclude high acceptance rates and an efficient use of resources, and allows providers to increase their revenues.
\end{abstract}


\end{frontmatter}


\input{01_introduction.tex}

\input{02_networkModel.tex}

\input{03_milpBkp.tex}

\input{04_evaluation.tex}

\input{05_relatedWork.tex}

\input{06_conclusion.tex}

\bibliography{references}

\end{document}

%% file: 01_introduction.tex
\section{Introduction}



Network virtualization has emerged as a powerful technique to allow multiple heterogeneous virtual networks to run over a shared infrastructure.
Production-level platforms recently proposed~\citep{Koponen2014,AlShabibi2014} give users the flexibility to run virtual networks with customized topologies and addressing schemes.
These solutions not only improve the operation of large enterprise infrastructures but also enable cloud operators to extend their service offerings of virtual storage and compute with network virtualization~\citep{Koponen2014}.

These modern platforms have two limitations that motivate our work.
First, they constrain the substrate to the infrastructure controlled by a single cloud operator.
This limits their scalability and dependability, as a single provider is effectively a single point of failure.
Several incidents in cloud facilities are evidence of this increasingly acute problem~\citep{Froehlich2015,Tsidulko2016} -- a problem that is today an important barrier to service adoption as more critical applications start shifting to the cloud.
This limitation motivates the exploration of availability-enhancing alternatives (\textit{e.g.}, through replication over multiple providers).

To overcome this problem, we have developed a platform that extends network virtualization to span multiple cloud providers~\citep{Alaluna17,Lacoste16}.
Enriching the substrate this way results in a number of benefits in terms of cost (e.g., by taking advantage of the pricing dynamics of cloud services~\citep{Zheng2015_2}), performance (for instance, by bringing services closer to their users), dependability, and security.
As an example of the latter, a tenant\footnote{We employ the terms user and tenant interchangeably in the paper.} that needs to comply with privacy legislation may request different parts of its virtual network to be placed in different locations. 
For instance, to abide by the recently implemented General Data Protection Regulation (GPDR), a virtual switch connecting databases that contain user data may need to be placed at a specific private location under the tenant's control, while the rest of the network is offloaded to public cloud infrastructures, to take advantage of their elasticity and flexibility.
With a multi-cloud substrate, an application can also be made immune to any single data center (or cloud availability zone) outage by spreading its services across providers. 

Existing solutions (e.g., VMware NSX~\citep{Koponen2014}) offer their tenants traditional networking services.
They allow tenants to specify a virtual topology of L2 switches and L3 routers, to define arbitrary addresses to their network elements, and to set ACL filtering rules.
Although this represents a formidable advance over the recent past (e.g., VLANs), it is still rather limited with respect to security and dependability.
Motivated by this limitation, we extend virtual networks with security assurances that go beyond simple ACL configurations.
Specifically, we allow users to specify security constraints for each element of the virtual network.
These constraints address, for instance, concerns about attacks on specific virtual hosts (\textit{e.g.}, covert channels or DDoS attacks) or on physical links (\textit{e.g.}, replay/eavesdropping).
In our solution, a tenant can define a high security level to particularly sensitive virtual nodes and/or links, mandating, respectively, their instantiation in secure VMs (e.g., an Amazon EC2 instance that provides DDoS protection), and their mapping to substrate paths that guarantee the required security properties (e.g., confidentiality).
To further extend the resiliency properties of our solution, and since we support the coexistence of resources from multiple clouds, both public and private, we assume each individual infrastructure (cloud) to have distinct levels of trust, from a user standpoint, enabling for instance virtual networks that are GPDR-compliant, as per the example above.

In this paper, we tackle the central resource allocation that is required to materialize a network virtualization solution -- the \emph{Virtual Network Embedding (VNE)} -- from this new perspective.
VNE addresses the problem of embedding the virtual networks specified by the tenants into the substrate infrastructure.
When a virtual network request arrives, the goal is to find an efficient mapping of the virtual nodes and links onto the substrate network, while maximizing the revenue of the virtualization provider.
This objective is subject to various constraints, such as the processing capacity on the substrate nodes and bandwidth of the links.

The literature on the problem is vast~\citep{surveyEmbedding}, but unfortunately, no existing solution meets all our requirements. 
In particular, the approaches that consider security are either limited in the level of protection offered, include assumptions that make them impractical for use in modern virtualization platforms, and/or do not fit an enriched multi-cloud substrate.
We further detail these differences in Section~\ref{sec:relatedWork}. 

Motivated by this gap in the literature, in this paper we propose an optimal VNE solution, based on Mixed Integer Linear Programming (MILP), that considers security constraints based on indications from the tenants.
These constraints address the concerns about attacks on hosts and on links mentioned above, including supporting the coexistence of resources (nodes/links) in multiple clouds with distinct levels of trust.
Given the limited expressiveness of a MILP formulation, we also propose a policy language to specify user requests.
Our language includes conjunction (\emph{`and'}), disjunction (\emph{`or'}), and negation (\emph{`not'}) operations, enabling tenants to express alternative resource requirements.

We have evaluated our proposal against the most commonly used VNE alternative~\citep{ViNEYard2012Boutaba}.
The results show that our solution makes a more efficient use of network resources, which is translated into higher acceptance ratios and reduced costs.
This demonstrates the advantage of our model for this context.
Another interesting takeaway is that the performance decrease is limited even when a reasonable number of virtual network requests includes security requirements.
For instance, when 20\% of requests include security demands, the reduction in acceptance ratio is of only 1 percentage point.
The results also illustrate the cost/revenue trade-off of including security services, 
shedding some light on the pricing schemes a virtualization operator should employ to benefit from offering these value-added services.

The contributions of our work can thus be summarized as:
(i) We formulate the SecVNE model and solve it as a Mixed Integer Linear Program (MILP). The novelty of our approach is in considering comprehensive security aspects over a multi-cloud deployment;
(ii) We propose a new policy language to specify the characteristics of the substrate network, and to allow the expression of user requirements;
(iii) We evaluate our formulation against the most commonly used VNE alternative~\citep{ViNEYard2012Boutaba}, and analyze its various trade-offs with respect to embedding efficiency, costs and revenues.



%% file: 02_networkModel.tex
\section{Multi-Cloud Network Virtualization}
\label{sec:networkModel}


Our multi-cloud network virtualization platform, Sirius, leverages from Software Defined Networking (SDN) to build a substrate infrastructure that spreads across both public clouds and private data centers~\citep{Alaluna17}.
These resources are then transparently shared by various tenants, allowing the definition and deployment of virtual networks (VN) composed of a number of virtual hosts (containers in our implementation) interconnected by virtual switches, arranged in an arbitrary network topology.
While specifying the virtual network, it is possible to indicate several requirements for the switches and links, for example with respect to the needed bandwidth, CPU capacity, and security guarantees.
These requirements are enforced during embedding by laying out the VN elements at the appropriate locations.
Specifically, those where the substrate infrastructure still has enough resources that fulfill the security requirements to satisfy the particular demands.
In addition, the datapaths are configured by the SDN controller by installing the necessary forwarding rules in the switches.

\begin{figure}
\begin{center}
\includegraphics[width=\textwidth]{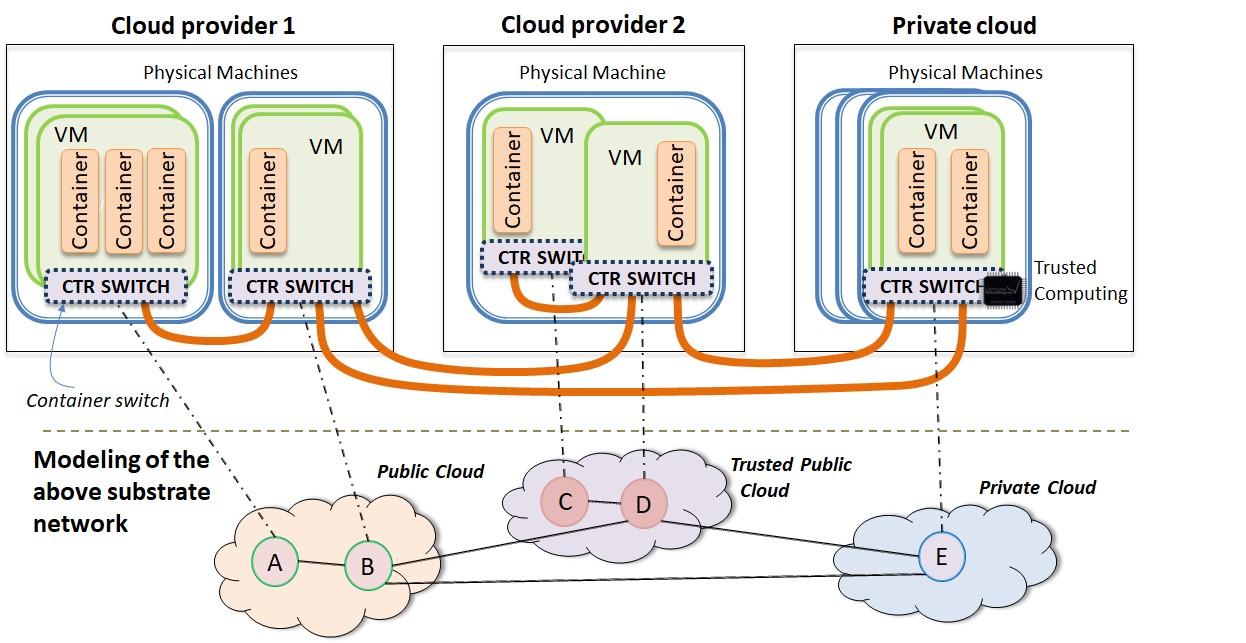}
\end{center}
\caption{Example substrate network encompassing resources from multiple clouds. }
\label{fig:substrate}
\vspace{-0.2cm}
\end{figure}

For example, as illustrated in Figure~\ref{fig:substrate}, virtual machines (VM) might be acquired by the virtualization operator at specific cloud providers to run tenants' containers implementing their distributed services.
In this scenario, the most relevant security aspects that may need to be assessed are the following.
First, the trust level associated with a cloud provider is influenced by various factors, which may have to be taken into consideration when running critical applications.
Providers are normally better regarded if they show a good past track record on breaches and failures, have been on the market for a while, and advertise Service Level Agreements (SLAs) with stronger assurances for the users. Moreover, as the virtualization operator has full control over its own data centers, it might employ protection features and procedures to make them compliant with regulations that have to be fulfilled by tenants (e.g., the EU GDPR that has recently come into force). 

Second, VMs can be configured with a mix of defense mechanisms, e.g., firewalls and antivirus, to build execution environments with stronger degrees of security at a premium price. These mechanisms can be selected by the operator when setting up the VMs, eventually based on the particular requirements of a group of tenants, or they could be sold ready to use by the cloud providers (e.g., like in Amazon\footnote{For example: Trend Micro Deep Security at aws.amazon.com/marketplace.} or Azure\footnote{For example: Check Point vSEC at azuremarketplace.microsoft.com.} offerings). Highly protected VMs arguably give more trustworthy conditions for the execution of the switch employed by the container manager, ensuring correct packet forwarding among the containers and the external network (e.g., without being eavesdropped or tampered with by malicious co-located containers). 

Third, the switches can also be configured with various defenses to protect network traffic.
In particular, it is possible to setup tunnels between switches implementing alternative security measures. For instance, if confidentiality is not a concern, then it is possible to add message authentication codes (MAC) to packets to afford integrity but without paying the full performance cost of encryption. Further countermeasures could also be added, such as denial of service detection and deep packet inspection to selected flows.
In some cases, if the trusted hardware is accessible (such as Intel CPUs with SGX extensions in the private cloud), one could leverage from it to enforce greater isolation while performing the cryptographic operations -- for example, to guarantee that secret keys are never exposed.

The reader should notice that the above discussion would also apply to other deployment scenarios.
For example, the virtualization operator could offer VNs of virtual machines (instead of containers) in distinct cloud providers.
This would require the use of nested virtualization mechanisms (e.g.,~\cite{Ben-Yehuda2010}), and in this case, the relevant network appliances would be the nested hypervisor switches and their corresponding inter-connections.
We have instead opted for using lightweight virtualization mechanisms, aligned with the increasing trend towards the use of microservices and other forms of serverless computing~\cite{vadhat2017}.\\

{\setlength{\parindent}{0cm}
\par{ {\bf Substrate Network Modeling.}}
Given the envisioned scenarios, the substrate network is modeled as a weighted undirected graph $ G^{S} = (N^{S}, E^{S}, A_{N}^{S}, A_{E}^{S}) $, composed of a set of nodes  $ N^{S} $ (e.g., switches/routers) and edges $ E^{S} $ interconnecting them.
Both the nodes and the edges have attributes that reflect their particular characteristics.
The collection of attributes we specify in our model resulted from conversations with several companies from the healthcare and energy sectors that are moving their critical services to the cloud, and they represent a balance among three goals: they should be
(i) expressive enough to represent the main security requirements when deploying virtual networks; 
(ii) easy to specify when configuring a network, requiring a limited number of options; 
and (iii) be readily implementable with available technologies. 
}

The following attributes are considered for substrate nodes:

\begin{center}$ A_{N}^{S} = \{ \{ cpu^{S} (n),~sec^{S} (n),~cloud^{S}(n)\}~|~n \in  N^{S}\} $ \end{center} 

The total amount of CPU that can be allocated for the switching operations of node $n$ is given by $ cpu^{S} (n) > 0 $.
Depending on the underlying machine capacity and the division of CPU cycles among the various tasks (e.g., tenant jobs, storage, network), $cpu^{S} (n)$ can take a greater or smaller value. The security level associated to the node is $ sec^{S} (n) > 0$. Nodes that run in an environment that implements stronger protections will have a  greater value for $ sec^{S} (n)$. The trustworthiness degree associated with a cloud provider is indicated with  $ cloud^{S}(n) > 0$.  
    
    The substrate edges have the following attributes: 
    
    \begin{center} $A_{E}^{S} = \{ \{ bw^{S} (l),~sec^{S} (l) \}~|~l \in E^{S} \} $ \end{center} 

The first attribute, $ bw^{S} (l) > 0$, corresponds to the total amount of bandwidth capacity of the substrate link \textit{l}. The security measures enforced by the link are reflected in $ sec^{S} (l) > 0$. If the link implements tunnels that ensure integrity and confidentiality (by resorting to MACs and encryption) then it will have a higher $ sec^{S} (l) $ than a default edge that simply forwards packets.\\


{\setlength{\parindent}{0cm}
\par{ {\bf Virtual Network Modeling.}} VNs have an arbitrary topology and are composed of a number of nodes and the edges that interconnect them. 
When a tenant wants to instantiate a VN, besides indicating the nodes' required processing capacity and bandwidth for the links’, she/he may also include as requirements security demands. These demands are defined by specifying security attribute values associated with the resources. 
}

In terms of modeling, a VN is also modeled as a weighted undirected graph, $ G^{V} = (N^{V},~E^{V},~A_{N}^{V},~A_{E}^{V}) $, composed of a set of nodes  $ N^{V} $ and edges (or links) $ E^{V} $.
Both the nodes and the edges have attributes that portray characteristics that need to be fulfilled when embedding is performed. Both $A_{N}^{V}$ and $A_{E}^{V}$ mimic the attributes presented for the substrate network.
The only exception is an additional attribute that allows for the specification of security requirements related to availability.

This attribute, $ avail^{V}(n) $, indicates that a particular node should have a backup replica to be used as a cold spare.
This causes the embedding procedure to allocate an additional node and the necessary links to interconnect it to its neighboring nodes.
These resources will only be used in case the virtualization platform detects a failure in the primary (or working) node.
$ avail^{V}(n)$ defines where the backup of virtual node \textit{n} should be mapped.
If no backup is necessary, $ avail^{V}(n) $ = 0.
If virtual node \textit{n} requires a backup to be placed in the same cloud, then $ avail^{V}(n) $ = 1.
Otherwise, if \textit{n} should have a backup in another cloud (e.g., to survive cloud outages), then $ avail^{V}(n) $ = 2.



\section{Secure Virtual Network Embedding Problem}
\label{sec:secVNE}

Our approach to VNE enables the specification of VNs to be mapped over a multi-cloud substrate, enhancing the security and flexibility of network virtualization.
More precisely, we define the Secure Virtual Network Embedding (SecVNE) problem as follows:

\vspace{0.3cm}
\noindent {\bf SecVNE problem:} \emph{Given a virtual network request (VNR) with resources and security demands $ G^{V}$, and a substrate network $ G^{S} $ with the resources to serve incoming VNRs, can $ G^{V} $ be mapped to $ G^{S} $ ensuring an efficient use of resources while satisfying the following constraints? (i) Each virtual edge is mapped to the substrate network meeting the bandwidth and security constraints; (ii) Each virtual node is mapped to the substrate network meeting the CPU capacity and security constraints (including node availability and cloud trust domain requirements).}
\vspace{0.3cm}


Our approach handles the SecVNE problem, mapping a VN onto the substrate network respecting all constraints.
When a VNR arrives, the goal of the embedding solution is to minimize the cost of mapping, while fulfilling all requirements.
In the situations when there are not enough substrate resources available, the incoming request is rejected.
To minimize these events, and therefore increase the VN acceptance ratio, we may allocate resources that are of a higher level of security than the ones specified in the VNR.
We find this option to represent a good trade-off.
For typical workloads, the alternative of having a more strict mapping will reduce the acceptance ratio while causing resources to be underutilized.

\begin{figure}
\begin{center}
\includegraphics[width=\textwidth]{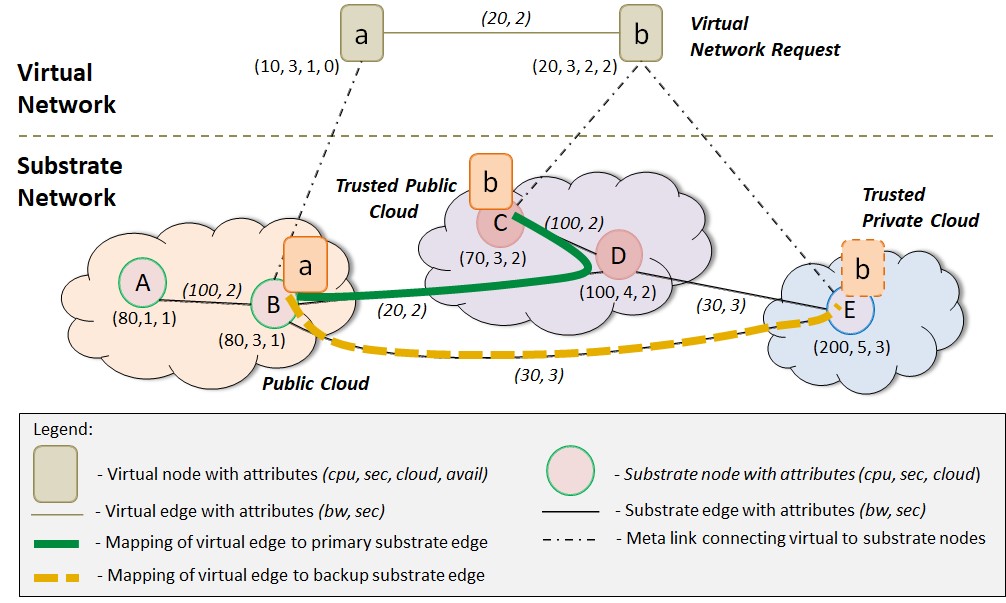}
\end{center}
\caption{Example of the embedding of a virtual network request (top) onto a multi-cloud substrate network (bottom). The figure also illustrates the various constraints and the resulting mapping after the execution of our MILP formulation.}
\label{fig:exampleEmbedding}
\vspace{-0.2cm}
\end{figure}


Figure \ref{fig:exampleEmbedding} shows an example of embedding a VNR (displayed on top) onto the substrate of Figure~\ref{fig:substrate} (represented at the bottom).
We assume a maximum security level of 5 for virtual nodes, and of 3 for virtual links.
We also assume in this example a cloud trustworthiness of 1 for the public cloud, of 2 for the trusted public cloud, and of 3 for the trusted private cloud.
These values were chosen arbitrarily for the sake of this example -- the operator will set these parameters to fit the specificities of its substrate.
The VNR consists of two nodes interconnected by one virtual link.
Node $a$ requires 10 units of CPU ($cpu^{V}(a) = 10$), a medium level of security ($sec^{V} (a) = 3$), and a default cloud trust level ($cloud^{V}(a) = 1$).
Besides slightly different CPU and security demands, the user requires the second node to be replicated.
Further, the primary and backup nodes should be placed in different clouds ($avail^{V} (b) = 2$).
For the virtual link the user requires 20 units of bandwidth ($bw^{V} (a,b) = 20$), and a medium level of security ($sec^{V} (a,b) = 2$).

The resulting embedding guarantees that all requirements are satisfied.
Specifically, node $a$ is mapped onto a substrate node with a security level equal to the one requested, on a public cloud ($sec^{S}(B) = 3$ and $cloud^{S}(B) = 1$).
The other virtual node and its backup are embedded on different clouds that fulfill the trustworthiness request (respectively, $cloud^{S}(C) = 2$ and $cloud^{S}(E) = 3$).
It is also possible to observe that one of the substrate paths (namely, the primary/working) maps to two substrate edges ($(B, D)$ and $(D, C)$), but guaranteeing that they both have the necessary security level ($2$, in this case).
In addition, the primary and backup paths are disjoint.
As will be justified later, this is a requirement of our solution.
The figure also displays meta-links connecting the virtual nodes to the substrate nodes they are mapped onto (e.g., the dash-dotted line between $a$ and $B$).
This is an important artifact in our modeling that is going to be explored in the MILP formulation (Section~\ref{sec:mipFormulationPartialBkp}).

\begin{figure}
\begin{center}
\includegraphics[width=\textwidth]{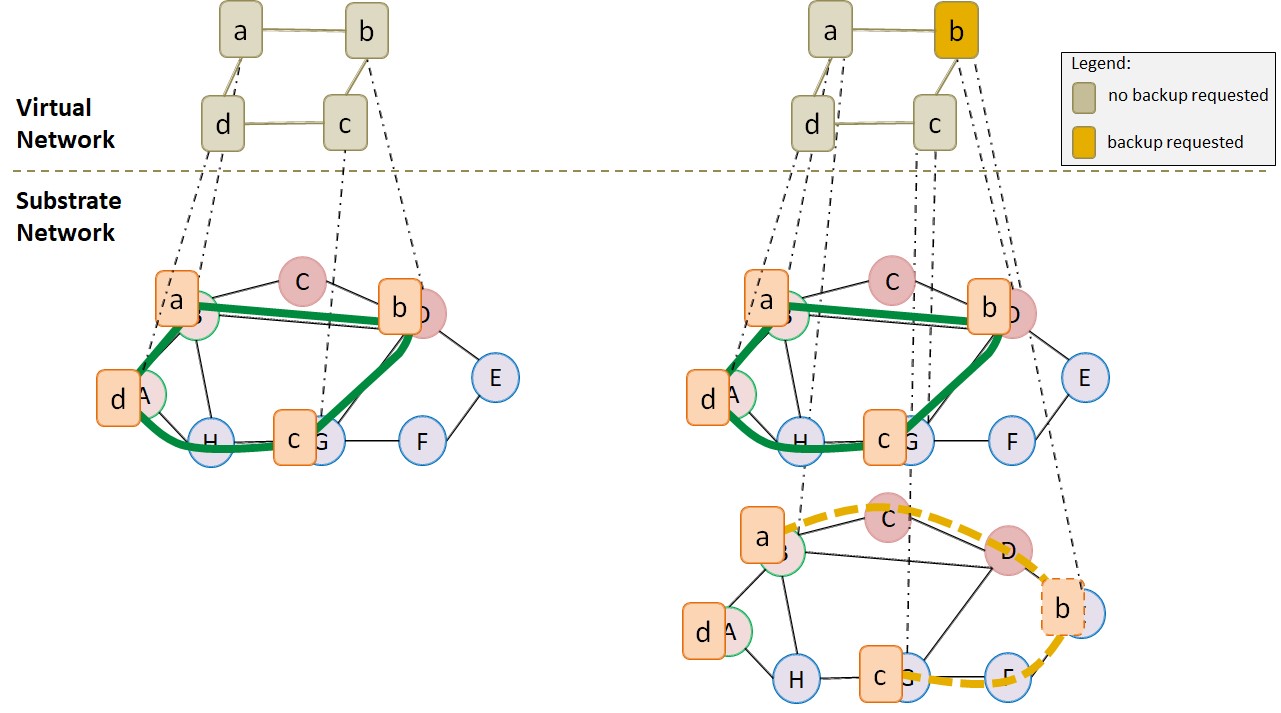}
\end{center}
\caption{Network model when no backup is requested (left); and when at least one backup node is requested (right).}
\label{fig:exampleBackupEmbedding}
\vspace{-0.2cm}
\end{figure}

Before closing this section, we detail how we model differently the VNRs that include backup requests, from those that do not have such requirements.
Please refer to Figure~\ref{fig:exampleBackupEmbedding}.
For VNRs where no node requests a backup, we only model the working network (Figure~\ref{fig:exampleBackupEmbedding}, left).
However, when at least one node requires backup, we model two networks (Figure~\ref{fig:exampleBackupEmbedding}, right): the working network, mapping all nodes and all primary paths; and the backup network.
The backup network also includes all nodes.
However, nodes that require backup are mapped to a \emph{different} substrate node from that of the working node, whereas nodes that do not require backup will be placed in the \emph{same} substrate node.
As this is an artifact of the model, no resources are reserved to this latter node.
The backup network includes only the backup paths that interconnect the backups to their neighbors.
As such, nodes that do not require backup and are not backup neighbors do not need to be connected to the rest of the network (one example is node $d$ in the figure).






\section{A Policy Language to Specify SecVNE} \label{sec:policy}

We support two alternative ways for both the virtualization operator and for the tenant to describe the substrate network and the virtual network, respectively.
The first is based on a graphical interface where the users can draw their arbitrary topologies, including nodes, links, and the associated attributes.
In this case, after the tenants draw their virtual network and launch a VNR, the solution runs the embedding solution we present in this paper (Section~\ref{sec:mipFormulationPartialBkp}) to map the request into the substrate.

As a result of our conversations with potential users of our platform (namely, enterprises that run critical applications), we have come to realize that a common requirement was that of expressing different options.
A typical example was that of a virtual node for which a high level of security is required if placed in a less trusted infrastructure (e.g., a public cloud), but that could have its security requirement lightened if placed in a highly trusted infrastructure (e.g., a private cloud).
Indeed, a problem with this first approach is that it does not grant tenants with this level of expressiveness.

We thus introduce a second approach based on a policy language, that lets the user describe both the substrate and the VNRs in a programmatic fashion.
The production rules of the grammar were kept relatively simple, but the achieved level of expressiveness is much greater than what is attained with the graphical interface alone.
As the characteristics of the substrate and VNRs are distinct, we explain them separately in the following.

\begin{table}[!h]
\centering
\begin{tabular}{ |p{9.2cm}| }
  		\hline
 		\textbf{Substrate Specification} \\
  		\hline
        \vspace*{0.002cm}
  		$S \rightarrow func^{S}(parameter) = value_{num}$  \\
        $S \rightarrow S\ \&\ S$  \\ 
        \hline \hline
 		\textbf{Virtual Network Specification} \\
        \hline 
       \vspace*{0.002cm}
        $V \rightarrow func^{V}(parameter) = value_{num}$  \\
        $V \rightarrow func^{V}(parameter) \geq value_{num}$  \\
        $V \rightarrow \ !V; \ (V);\ V\ \&\ V;\ V\ |\ V $  \\ 
        \hline
\end{tabular}
\caption{Policy grammar to define SecVNE parameters.}
\label{tab:policy}
\end{table}

The substrate part of the SecVNE policy grammar (top rows of Table~\ref{tab:policy}) enables the listing of resources that compose the substrate.
The parameters represent the substrate node or link ID, and the function enables the specification of particular attributes.
For example, the leftmost cloud of Figure~\ref{fig:exampleEmbedding} is specified as:\\

$substrate \rightarrow cpu^S(A)=80\ \&\ sec^S(A)=1\ \&\ cloud^S(A)=1\ \&\\
~~~~~~~~~~~~~~~~~~~~~~~ cpu^S(B)=80\ \&\ sec^S(B)=3\ \&\ cloud^S(B)=1\ \&\\
~~~~~~~~~~~~~~~~~~~~~~~ bw^S(A,B)=100\ \&\ sec^S(A,B)=2\ \&\ ...
$\\



In the virtual part of the SecVNE policy grammar (bottom rows of Table~\ref{tab:policy}), the relations dictate the requirements for each node and link of the VNR.
Differently from the substrate part, certain functions specify exact values, whereas others specify a minimum demand.
For instance, while the CPU requirement is an exact value, the security function specifies the minimum security level required, as a virtual node can be mapped to a substrate node with either the same or a higher security level.
As the grammar supports boolean operations, including $or$ (``$|$"), $and$ (``\&"), and $not$ (``!"), it is possible to express alternative resource requirements.

As an example, consider the following VN with two nodes and one edge:\\

$VN \rightarrow (CPU^V(a)=10\ \&\ sec^V(a)\geqslant3\ \&\ cloud^V(a)\geqslant1\ \&\ \\ ~~~~~~~~~~ avail^V(a)=0)\ \&\ (CPU^V(b)=20\ \&\ avail^V(b)=1\ \&\\
~~~~~~~~~~ ((sec^V(b)\geqslant1\ \&\ cloud^V(b)\geqslant4)\ |\  (sec^V(b)\geqslant4\ \&\ cloud^V(b)\geqslant1))\ \&\\
~~~~~~~~~~ (bw^V(a,b)=20\ \&\ sec^V(a,b)\geqslant2$)\\

Node $a$ requires a security level of at least $3$, and cloud trustworthiness of $1$ or higher.
For node $b$, on the other hand, the tenant makes a compromise between node security and the degree of cloud trust.
If the cloud has a high level of trust (at least $4$), then this node's security level can be as low as $1$. 
However, the tenant is willing to have this virtual node mapped in a less trusted cloud (with a trust level of only $1$), but in that case, the node security level is increased to at least $4$. 

When processing a VNR containing a VN with several optional demands as this one, we generate all possible requests that would satisfy the tenant.
Then, we evaluate each one and select the solution with the lowest cost.
There are two main benefits to this approach.
First, as explained, the increased expressiveness allows tenants to explore different trade-offs with respect to security and availability of their resources.
Second, by evaluating several solutions we increase the number of options, reducing mapping costs and, as a consequence, enabling higher acceptance ratios.


%% file: 03_milpBkp.tex
\section{MILP Formulation} \label{sec:mipFormulationPartialBkp}

In this section we present our MILP formulation to solve the SecVNE problem.
The section starts by explaining the decision variables used in the formulation, proceeds to present the objective function, and finally the constraints required to model the problem.

\begin{table}[t]
\footnotesize
\begin{tabular}{ |c| p{13.7cm}| }
  		\hline
 		\multirow{2}{*}{\textbf{Symbol}} & \multirow{2}{*}{\textbf{~~~~~~~~~~~~~~~~~~~~~~Meaning}}\\&  \\
  		\hline
  		\multirow{2}{*}{$wf_{p,q}^{i,j} \geqslant 0$} & \multirow{2}{*}{\parbox{13.7cm}{The amount of working flow, i.e., bandwidth, on physical link (\textit{p},\textit{q}) for virtual link (\textit{i},\textit{j})}}\\  & \\\hline
  		\multirow{2}{*}{$bf_{p,q}^{i,j} \geqslant 0$} & \multirow{2}{*}{\parbox{13.7cm}{The amount of backup flow, i.e., backup bandwidth, on physical link (\textit{p},\textit{q}) for virtual link (\textit{i},\textit{j})}} \\ & \\\hline
  		\multirow{2}{*}{$wl_{p,q}^{i,j} \in \{0,1\}$ } & \multirow{2}{*}{\parbox{13.7cm}{Denotes whether virtual link (\textit{i},\textit{j}) is mapped onto physical link (\textit{p},\textit{q}). (1 if (\textit{i},\textit{j}) is mapped on (\textit{p},\textit{q}), 0 otherwise)}}\\ & \\\hline
  		\multirow{3}{*}{$bl_{p,q}^{i,j} \in \{0,1\}$ } & \multirow{3}{*}{\parbox{13.7cm}{Denotes whether the backup of virtual link (\textit{i},\textit{j}) is mapped onto physical link (\textit{p},\textit{q}). (1 if backup of (\textit{i},\textit{j}) is mapped on (\textit{p},\textit{q}), 0 otherwise)}}\\ & \\ & \\\hline
  		\multirow{2}{*}{$wn_{i,p} \in \{0,1\}$ } & \multirow{2}{*}{\parbox{13.7cm}{Denotes whether virtual node \textit{i} is mapped onto physical node \textit{p}. (1 if \textit{i} is mapped on \textit{p}, 0 otherwise)}} \\ & \\\hline
  		\multirow{2}{*}{$bn_{i,p} \in \{0,1\}$} & \multirow{2}{*}{\parbox{13.7cm}{Denotes whether virtual node \textit{i}'s backup is mapped onto physical node \textit{p}. (1 if \textit{i}'s backup is mapped on \textit{p}, 0 otherwise)}} \\ & \\\hline
        \multirow{2}{*}{$wc _{i,c} \in \{0,1\}$} & \multirow{2}{*}{\parbox{13.7cm}{Denotes whether virtual node \textit{i} is mapped on cloud \textit{c}. (1 if \textit{i} is mapped on \textit{c}, 0 otherwise)}} \\ & \\\hline
        \multirow{2}{*}{$bc _{i,c} \in \{0,1\}$} & \multirow{2}{*}{\parbox{13.7cm}{Denotes whether virtual node \textit{i}'s backup is mapped on cloud \textit{c}. (1 if \textit{i}'s backup is mapped on \textit{c}, 0 otherwise)}} \\ & \\\hline
\end{tabular}
\caption[MILP formulation variables]{Domain constraints  (decision variables) used in the MILP formulation.}
\label{tab:variablesTab}
\end{table}


\subsection{Decision variables and auxiliary parameters}
\label{sec:decisonVar}

Table~\ref{tab:variablesTab} presents the variables that are used in our MILP formulation.
Briefly, $wf_{p,q}^{i,j}$, $bf_{p,q}^{i,j}$, $wl_{p,q}^{i,j}$ and $bl_{p,q}^{i,j}$ are related to (working and backup) links; 
$wn_{i,p}$ and $bn_{i,p}$ are associated with (working and backup) nodes; 
and $wc _{i,c}$ and $bc _{i,c}$ are related to the cloud location of a virtual node.
In Table~\ref{tab:auxVariablesTab} we present a few additional parameters used on the formulation.
Their importance will be made clear as we describe the solution.

\begin{table}[t]
\footnotesize
\begin{tabular}{ |c| p{13.4cm}| }
  		\hline
 		\multirow{2}{*}{\textbf{Symbol}} & \multirow{2}{*}{\textbf{~~~~~~~~~~~~~~~~~~~~~~Meaning}}\\&  \\
  		\hline
  		\multirow{2}{*}{$\beta_{1},~\beta_{2},~\beta_{3}$} & \multirow{2}{*}{\parbox{13.4cm}{Coefficients used in the \textit{Objective Function} to provide a weighted-sum properly parameterized for each objective.}}\\  & \\\hline
  		\multirow{2}{*}{$ \alpha_{p,q} $} & \multirow{2}{*}{\parbox{13.4cm}{A weight representing the relative cost of link \emph{(p,q)}.}}\\  & \\\hline
  		\multirow{2}{*}{$nodeLocation_{p,c}$} & \multirow{2}{*}{\parbox{13.4cm}{Denotes the location of substrate node $p$. (1 if substrate node $p$ is located in cloud $c$; 0 otherwise)}} \\ & \\\hline
  		\multirow{2}{*}{$backupNetwork$} & \multirow{2}{*}{\parbox{13.4cm}{Assumes value 1 if at least one of the nodes of a VNR requests backup; 0 otherwise.}}\\ & \\\hline
  		\multirow{2}{*}{$working_{p,q}$ } & \multirow{2}{*}{\parbox{13.4cm}{Auxiliary binary variable defining if a physical link $(p, q)$ is part of the working network.}}\\ & \\\hline
  		\multirow{2}{*}{$backup_{p,q}$ } & \multirow{2}{*}{\parbox{13.4cm}{Auxiliary binary variable defining if a physical link $(p, q)$ is part of the backup network.}} \\ & \\\hline
\end{tabular}
\caption[MILP additional parameters]{Additional parameters used in the MILP formulation.}
\label{tab:auxVariablesTab}
\end{table}

\begin{table}[h]
\centering
\footnotesize
\renewcommand*{\arraystretch}{1.3}
\begin{tabular}{ |l|  }
\hline
$\mathring{N}^V$ = $\{~ i \in N^V~:~avail^V(i) = 0~\}$\\ 
$\overline{N}^V$ = $N^V \setminus \mathring{N}^V$\\ 
$\mathring{E}^V$ = $\{~ (i,j) \in E^V~:~avail^V(i) = 0~$and$~avail^V(j) =0~\}$\\
$\overline{E}^V$ = $E^V \setminus \mathring{E}^V$
\\\hline
\end{tabular}
\caption{Auxiliary sets to facilitate the description of the formulation constraints.}
\label{tab:sets}
\end{table}

The formulation also employs a few auxiliary sets whose value depends on the VNR, as shown in Table~\ref{tab:sets}.
For example, $\mathring{N}^V$ is a set that includes all nodes that do not require backup, and $\overline{N}^V$ is its complement. 
As such, when $ \overline{N}^V = \emptyset$ that means no virtual node requires backup.
We recall that, when this happens, we only model a working network.
This means that every backup-related decision variable  ($bf_{p,q}^{i,j}$, $bl_{p,q}^{i,j}$, $bn_{i,p}$, $bc _{i,c}$) is equal to 0.
On the other hand, if $ \overline{N}^V \neq \emptyset$, then we model both a working and a backup network (recall Figure~\ref{fig:exampleBackupEmbedding}).
When we need to model a backup network (i.e., when at least one node requests backup), then we introduce another artifact into our model.
Namely, if virtual node $i$ has $avail^{V}(i) = 0$, indicating that it does not require replication, then both the working and the backup nodes of $i$ are placed in the same substrate node $p$ (i.e., $wn_{i,p}=bn_{i,p}$ $= 1$).
However, it is guaranteed that this second node, the ``virtual'' (i.e., not requested) backup, does not consume resources (e.g., CPU).
When a virtual node $j$ has $avail^{V}(j) > 0$, thus requiring replication, it is necessary to map the working and backup in different substrate nodes (possibly in distinct clouds, as explained in Section~\ref{sec:networkModel}).
In this case, the backup will have the necessary resources reserved, to be able to substitute the primary in case of failure.

\subsection{Objective Function} \label{sec:objFunction}

The objective function aims to minimize three aspects (see Eq. \ref{eq:obj}):
1) the sum of all computing costs,
2) the sum of all communication costs,
and 3) the overall number of hops of the substrate paths used to map the virtual links.
Since these objectives are measured in different units, we resort to a composite function, which can be parametrized and used to compute different solutions (others approaches could be used, see Steuer~\citep{steuer86a}).
Thus, the formulation is based on a weighted-sum function with three different coefficients, $\beta_{1}$, $\beta_{2}$, and $\beta_{3}$, which should be reasonably parameterized by the virtualization operator for each objective.

\begin{align}
		min \qquad &
        \beta_{1} \Big[ \smashoperator[r]{\sum_{i\in N^{V}}} ~~
		    \smashoperator[r]{\sum_{p\in N^{S}}} ~
		    cpu^{V}(i) \quad sec^{S}(p) \quad cloud^{S}(p) \quad wn_{i,p}\nonumber \\
        &
        + \smashoperator[r]{\sum_{i\in \overline{N}^{V}}} ~~
		    \smashoperator[r]{\sum_{p\in N^{S}}} ~
		    cpu^{V}(i) \quad sec^{S}(p) \quad cloud^{S}(p) \quad bn_{i,p} \Big] \nonumber \\
		&
        + \beta_{2} \Big[ \smashoperator[r]{\sum_{(i,j)\in E^{V}}} ~~
		    \smashoperator[r]{\sum_{(p,q) \in E^{S}}} ~ 
		    \alpha_{p,q} \quad sec^{S}(p,q) \quad wf_{p,q}^{i,j} \nonumber \\
            &
		+ \smashoperator[r]{\sum_{(i,j)\in \overline{E}^{V}}} ~~
		    \smashoperator[r]{\sum_{(p,q) \in E^{S}}}~\alpha_{p,q} \quad 
            sec^{S}(p,q) \quad bf_{p,q}^{i,j} \Big] \nonumber \\
		& 
        +  \beta_{3} \Big[ \sum_{(i,j) \in E^{V}} \sum_{(p,q) \in E^{S}}  wl_{p,q}^{i,j} 
        ~+  \sum_{(i,j) \in \overline{E}^{V}} \sum_{(p,q) \in E^{S}} ~bl_{p,q}^{i,j} \Big] \nonumber \\
	 	\label{eq:obj}
\end{align}


The first part of Eq.~\ref{eq:obj} covers the computing costs, including both the working and backup nodes (top 2 lines). 
The second part of the equation is the sum of all working and backup link bandwidth costs (lines 3-4).
The last part of the objective function is related to the number of hops of the working and backup paths.

A few aspects are worth detailing.
First, note that for the node backup case we restrict the sum to the set of nodes that require backup ($\overline{N}^V$).
This is needed to express that the nodes that do not require backup (the ``virtual'' backups referred above) do not consume resources.
The equation considers the level of security of the substrate resources and the trustworthiness of the cloud infrastructure that hosts them.
This is based on the assumption that a higher level of security or of trustworthiness is translated into an increased cost.
As the parameters that formalize these levels ($sec^{S}(p)$ and $cloud^{S}(p)$) can take as value any positive real number, this allows the virtualization operator to fine-tune its costs. 
To address the possibility that different substrate edges may have different costs, we have included a multiplicative parameter $ \alpha_{p,q} $ in lines 3-4.
This parameter is a weight that can be used to express these differences.
For instance, a link connecting two clouds might have a higher (monetary, delay, or other) cost, when compared to intra-cloud links.

Intuitively, this objective function attempts to economize the most ``powerful'' resources (e.g., those with higher security levels) for VNRs that explicitly require them. Therefore, for instance, it is expected in most cases virtual nodes with $sec^{V} = 1$ to be mapped onto substrate nodes with $sec^{S} = 2$  only if there are no other substrate nodes with $sec^{S} = 1$ available.

\subsection{Security Constraints}

Next, we enumerate the constraints related to the security of (working and backup) nodes, edges, and clouds, respectively:

\begin{align}	
	wn_{i,p} ~~ sec^{V}(i) \leqslant sec^{S}(p),~\forall i\in N^{V},~p\in N^{S} \label{eq:wvnodesec}
    \\
	bn_{i,p} ~~ sec^{V}(i) \leqslant sec^{S}(p),~\forall i\in N^{V},~p\in N^{S} \label{eq:bsnodesec}
    \\
    wl_{p,q}^{i,j} ~~ sec^{V}(i,j) \leqslant sec^{S}(p,q),~\forall (i,j)\in E^{V},~(p,q)\in E^{S} \label{eq:wvlinksec} 
	\\
	bl_{p,q}^{i,j} ~~ sec^{V}(i,j) \leqslant sec^{S}(p,q),~\forall (i,j)\in E^{V},~(p,q)\in E^{S} \label{eq:bvlinksec}
    \\
    wn_{i,p} ~~ cloud^{V}(i) \leqslant cloud^{S}(p),~\forall i \in N^{V},~p \in N^{S} \label{eq:wcloud}
	\\
	bn_{i,p} ~~ cloud^{V}(i) \leqslant cloud^{S}(p),~\forall i \in N^{V},~p \in N^{S} \label{eq:bcloud}
\end{align}

The first of these constraints guarantees that a virtual node is only mapped to a substrate node that has a security level that is equal to or greater than its demand (Eq.~\ref{eq:wvnodesec}).
The next equation guarantees the same for backup nodes.
The following two equations force each virtual edge to be mapped to (one or more) physical links that provide a level of security that is at least as high as the one requested.
This is guaranteed for links connecting the primary nodes (Eq.~\ref{eq:wvlinksec}) and the backups (Eq.~\ref{eq:bvlinksec}). 
The last constraints ensure that a virtual node $i$ is mapped to a substrate node $p$ only if the cloud where $p$ is hosted has a trust level that is equal or greater than the one demanded by $i$ (Eq. \ref{eq:wcloud} and \ref{eq:bcloud} for working and backup, respectively).

\subsection{Mapping Constraints}

{\setlength{\parindent}{0cm}
We take security, including availability, as a first class citizen in our solution.
Therefore, when faced with a choice between security and resource efficiency, we give preference to the former, as will be made clear next.
\par{ {\it Node Embedding:}} 
We force each virtual node to be mapped to exactly one working substrate node (Eq.~\ref{eq:wnodemapp}) and, when a backup is requested, to a single backup substrate node (Eq.~\ref{eq:bnodemapp}).
We also guarantee that a substrate node maps at most one virtual node (working or backup) of a single tenant.
This means we have opted to avoid substrate sharing in a tenant's virtual network, to improve its availability.
As a result, if one substrate node fails, this will affect at most one (backup or working) virtual node.
This is one example of our design choice of availability over efficiency.
This is expressed in three equations -- Eq. ~\ref{eq:mappN1} to \ref{eq:mappN2} -- due to the use of the ``virtual'' backup artifact in our model.
Eq.~\ref{eq:mappN1} guarantees that one substrate node maps at most one working virtual node.
The next equation guarantees the same for the backup case.
Eq.~\ref{eq:mappN2} guarantees that one substrate node will not map both a backup node and a virtual node.
Note the use of the set $\overline{N}^{V}$ to guarantee that the ``virtual'' backups are not included -- they are the exception to the rule.
As explained in Section~\ref{sec:decisonVar}, if a virtual node requires no replication but there is a backup network (as at least one other node has requested it), its working and backup are mapped onto the same substrate node. 
This is formalized with the next two equations.
Note they both use the set $\mathring{N}^{V}$, meaning that they deal only with virtual nodes that do not require backup.
Eq. \ref{eq:mappN4} guarantees that ``virtual'' backup nodes can only be mapped to their corresponding working node, while Eq. \ref{eq:mappN3} grants these substrate nodes the exception of hosting a maximum of 2 virtual nodes: the working and the ``virtual'' backup node.    
}
\begin{align}
	\smashoperator{\sum_{p \in N^{S}}} wn_{i,p} = 1 ,~\forall i \in N^{V} \label{eq:wnodemapp}\\
	\smashoperator[r]{\sum_{p \in N^{S}}} bn_{i,p} = 1 ,~\forall i \in \overline{N}^{V} \label{eq:bnodemapp}\\
    \smashoperator[r]{\sum_{i\in N^{V}}} wn_{i,p} \leqslant 1 ,~\forall p \in N^{S} \label{eq:mappN1}\\
    \smashoperator[r]{\sum_{i\in N^{V}}} bn_{i,p} \leqslant 1 ,~\forall p \in N^{S} \label{eq:mapb}\\
    \smashoperator[r]{\sum_{i\in N^{V}}} wn_{i,p} + bn_{j,p} \leqslant 1,~\forall j \in \overline{N}^{V},~p \in N^{S} \label{eq:mappN2}\\
    bn_{i,p} \leqslant wn_{i,p},~\forall i \in \mathring{N}^{V},~p \in N^{S} \label{eq:mappN4}\\
    \smashoperator[r]{\sum_{i\in N^{V}}} wn_{i,p} + bn_{j,p} \leqslant 2,~\forall j \in \mathring{N}^{V},~p \in N^{S} \label{eq:mappN3}
\end{align}



The next set of constraints create the necessary relationships between nodes and the flows that traverse them.

\begin{align}
    wl^{i,j}_{p,q}~~bw^{V}(i,j) \geqslant wf^{i,j}_{p,q},~\forall (i,j) \in E^{V},~(p,q) \in E^{S} \label{eq:wlwfrelation}\\
bl^{i,j}_{p,q}~~bw^{V}(i,j) \geqslant bf^{i,j}_{p,q},~\forall (i,j) \in E^{V},~(p,q) \in E^{S} \label{eq:blbfrelation}\\
	wl^{i,j}_{p,q} = wl^{i,j}_{q,p},~\forall (i,j) \in E^{V},~p,q \in N^{S} \cup N^{V} \label{eq:bincon1}
	\\
	bl^{i,j}_{p,q} = bl^{i,j}_{q,p},~\forall (i,j) \in E^{V},~p,q \in N^{S} \cup N^{V} \label{eq:bincon2}
\end{align}

Eq.~\ref{eq:wlwfrelation} ensures that if there is a flow between nodes $p$ and $q$ for a virtual edge $(i,j)$, then $(i,j)$ is mapped to the substrate link whose end-points are $p$ and $q$.
The use of the inequality in this equation is important, to guarantee the possibility of network flows to use multiple paths. 
The next equation achieves the same goal, but for the backup.
We also include two binary constraints to force the symmetric property for the binary variables that define the link mappings (Eqs. \ref{eq:bincon1} and \ref{eq:bincon2}).
Note that in these equations we included links from both the substrate and the virtual networks.
This is due to the need to include the meta-links described in Section~\ref{sec:secVNE}.

Similarly, we also need to establish the relation between the virtual nodes and the clouds they are embedded into.
This is achieved with Eqs. \ref{eq:cloudsubvir} and \ref{eq:cloudsubvir2}, for working and backups, respectively.
Specifically, if virtual node \textit{i} is mapped onto a substrate node \textit{p}, and \textit{p} is hosted in cloud \textit{c}, then \textit{i} is mapped into cloud \textit{c}.
The auxiliary parameter $nodeLocation_{p,c}$ has value 1 if substrate node $p$ is hosted in cloud $c$, and 0 otherwise.
We also require each virtual node to be mapped to exactly one cloud (working or backup), with Eqs. \ref{eq:wncloud} and \ref{eq:bncloud}.
The auxiliary parameter $backupNetwork$ assumes value 1 if a backup is needed for at least one of the nodes of a VNR, or value 0 otherwise.

\begin{align}
    	\sum_{p \in N^{S}} (wn_{i,p}~~nodeLocation_{p,c}) \geqslant wc_{i,c},~\forall i \in N^{V},~c \in C \label{eq:cloudsubvir} \\
    \sum_{p \in N_{S}} (bn_{i,p}~~nodeLocation_{p,c}) \geqslant bc_{i,c},~\forall i \in N^{V},~c \in C \label{eq:cloudsubvir2}\\
    \sum_{c \in C} wc_{i,c} = 1,~\forall i \in N^{V} \label{eq:wncloud}\\
\sum_{c \in C} bc_{i,c} = backupNetwork,~\forall i \in N^{V}\label{eq:bncloud}
\end{align} 

We give tenants three replication options for their virtual nodes: no replication, replication in the same cloud, and replication in a different cloud.  
We must thus restrict the placement of the working and backup nodes to the same or to distinct clouds, depending on the value of the availability attribute ($avail^{V}(i)$).
This is achieved with Eq. \ref{eq:cLoc1}\footnote{Note that in the implementation the \textit{modulus} function was linearized.}.
In this equation, when $avail^{V}(i)=1$, variable $bc_{i,c}$ is equal to $wc_{i,c}$: the backup is mapped to the same cloud as the working node, as required.
By contrast, when $avail^{V}(i)=2$, $bc_{i,c}$ has to be different from $wc_{i,c}$, so the nodes will be mapped to different clouds.
Finally, when $avail^{V}(i)=0$, $bc_{i,c}$ will be equal to zero (this condition needs to be considered jointly with Eq.~\ref{eq:wncloud}).

\begin{align}
	|wc_{i,c}~~backupNetwork~-~bc_{i,c}| = ( avail^{V}(i) - 1)~\times\nonumber\\ (wc_{i,c}~~backupNetwork + bc_{i,c}),~ \forall i \in N^{V},~c \in C \label{eq:cLoc1}
\end{align}

{\setlength{\parindent}{0cm}
\par{ {\it Link Embedding:}} The next constraints are related to the mapping of virtual links into the substrate.
They take advantage of the meta link artifact (recall Figure~\ref{fig:exampleEmbedding}), which connects a virtual node $i$ to the substrate node $p$ where it is mapped, to enforce a few restrictions.
}
\normalsize
\begin{align}
     wn_{i,p} \quad bw^{V}(i,j) = wf_{i,p}^{i,j},~\forall ~(i,j) \in E^{V},~p\in N^{S} \label{eq:working1} \\
	wn_{j,q} \quad bw^{V}(i,j) = wf_{q,j}^{i,j},~\forall (i,j) \in E^{V},~q\in N^{S} \label{eq:working2}\\
	bn_{i,p} ~~ bw^{V}(i,j) = bf_{i,p}^{i,j} ~~ backupNetwork,~\forall (i,j) \in E^{V},~p\in N^{S} \label{eq:fail1}\\
	bn_{j,q} ~~ bw^{V}(i,j) = bf_{q,j}^{i,j} ~~ backupNetwork,~\forall (i,j) \in E^{V},~q\in N^{S} \label{eq:fail2}\\
    \smashoperator{\sum_{j, k != i \:\,\, j,k \in N^{V}}} wf_{i,p}^{j,k} + wf_{p,i}^{j,k} = 0,~ \forall i \in N^{V},~p\in N^{S} \label{eq:avoid}\\
\smashoperator{\sum_{j, k != i \:\,\, j,k \in N^{V}}} bf_{i,p}^{j,k} + bf_{p,i}^{j,k} = 0 ,~ \forall i \in N^{V},~p\in N^{S} \label{eq:avoidb}
\end{align}

These constraints guarantee that the working flow of a virtual link $(i,j)$ has its source in \textit{i} and its sink in \textit{j}, traversing the corresponding substrate nodes ($p$ and $q$) (Eq. \ref{eq:working1} and \ref{eq:working2}).
These equations effectively define the meta-link artifact.
The next two equations formalize the same requirement for the backup nodes.
Please note that even though the backup path is only used if the working substrate path fails, we reserve the necessary resources during embedding to make sure they are available when needed.
Eqs.~\ref{eq:avoid} and~\ref{eq:avoidb} force meta-links to carry only (working or backup, respectively) traffic to/from their virtual nodes.

The next equations specify flow conservation restrictions at the nodes. 

\normalsize
\begin{align}
	\smashoperator[r]{\sum_{p\in N^{S}}} wf_{i,p}^{i,j}~- \smashoperator[r]{\sum_{p\in N^{S}}} wf_{p,i}^{i,j} = bw^{V}(i,j),~\forall (i,j) \in E^{V} \label{eq:flows}\\
	\smashoperator[r]{\sum_{p\in N^{S}}} wf_{j,p}^{i,j}~- \smashoperator[r]{\sum_{p\in N^{S}}} wf_{p,j}^{i,j} = -bw^{V}(i,j),~\forall (i,j) \in E^{V} \label{eq:flowt}\\
	\smashoperator[r]{\sum_{p\in N^{S}\cup N^{V}}} wf_{q,p}^{i,j}~- \smashoperator[r]{\sum_{p\in N^{S}\cup N^{V}}} wf_{p,q}^{i,j} = 0,~\forall (i,j) \in E^{V} ,~q \in N^{S} \label{eq:flowcon} \\
  	\smashoperator[r]{\sum_{p\in N^{S}}}~bf_{i,p}^{i,j}~- \smashoperator[r]{\sum_{p\in N^{S}}}~bf_{p,i}^{i,j} = bw^{V}(i,j) ~~ backupNetwork,~\forall (i,j) \in E^{V} \label{eq:bflows}\\
	\smashoperator[r]{\sum_{q\in N^{S}}}~bf_{j,q}^{i,j}~- \smashoperator[r]{\sum_{q\in N^{S}}}~bf_{q,j}^{i,j} = -bw^{V}(i,j) ~~ backupNetwork,~\forall (i,j) \in E^{V} \label{eq:bflowt}\\
\smashoperator[r]{\sum_{p\in N^{S}\cup N^{V}}}~bf_{q,p}^{i,j}~- \smashoperator[r]{\sum_{p\in N^{S}\cup N^{V}}}~bf_{p,q}^{i,j} = 0,~ \forall (i,j) \in E^{V},~q \in N^{S} \label{eq:bflowcon}
\end{align}  

Eqs.~\ref{eq:flows}, \ref{eq:flowt} and \ref{eq:flowcon} refer to the working flow conservation conditions, which denote that the network flow to a node is zero, except for the source and the sink nodes, respectively. 
In an analogous way, the following three equations refer to the backup flow conservation conditions (Eq. \ref{eq:bflows},  \ref{eq:bflowt} and \ref{eq:bflowcon}).

The next set of constraints guarantee flow symmetry, to ensure the same flow traverses both directions, for both working (Eq. \ref{eq:wflowconeq}) and backup nodes (Eq.~\ref{eq:bflowconeq}).
Finally, Eqs. \ref{eq:wbflowcoeqbkp} and \ref{eq:wbflowcoeqbkp2} define the same flow for the working node and its backup, for both directions.

\normalsize
\begin{align}
wf_{p,q}^{i,j}~=wf_{q,p}^{j,i},~\forall (i,j) \in E^{V},~p,q \in N^{S} \cup N^V \label{eq:wflowconeq} \\
bf_{p,q}^{i,j}~=bf_{q,p}^{j,i},~\forall (i,j) \in E^{V},~p,q \in N^{S} \cup N^V\label{eq:bflowconeq}\\
wf_{p,q}^{i,j}~=bf_{p,q}^{i,j},~\forall (i,j) \in \mathring{E}^V,~p,q \in N^{S}\label{eq:wbflowcoeqbkp}\\
wf_{p,q}^{j,i}~=bf_{p,q}^{j,i},~\forall (i,j) \in \mathring{E}^V,~p,q \in N^{S}\label{eq:wbflowcoeqbkp2}
\end{align}

{\setlength{\parindent}{0cm}
\par{ {\it Nodes and Links Disjointness:}}
When a virtual node stops responding, it may have been the case that the node has effectively faulted, or the failure has occurred in one of the substrate links that guarantees its connectivity to other nodes.
Aligned with our goal of providing high guarantees of availability, we aim to cover these two cases.
For this purpose, we ensure that the paths connecting the virtual nodes' backup are disjoint from the substrate resources used for the working part (otherwise, a single failure could compromise both paths).
For this purpose, we introduce the auxiliary binary variables $working_{p,q}$ and $backup_{p,q}$, that define if a physical link $(p,q)$ belongs to the working or to the backup networks, respectively.
}
\begin{align}
    working_{p,q}  \leqslant 1 - backup_{p,q},~\forall (p,q) \in E^{S} \label{eq:wbDisj}\\
    wl^{i,j}_{p,q} \leqslant working_{p,q},~\forall (i,j) \in E^V,~(p,q) \in E^{S} \label{eq:wDisj2} \\
    bl^{i,j}_{p,q} \leqslant backup_{p,q},~\forall (i,j) \in \overline{E}^{V},~(p,q) \in E^{S} \label{eq:bDisj}
\end{align}

First, we require disjointness between the working and backup paths (Eq. \ref{eq:wbDisj}).
Second, we guarantee that, if the working path of a virtual edge $(i,j)$ is mapped onto a substrate link $(p,q)$, then $(p,q)$ is placed into the working network (Eq. \ref{eq:wDisj2}).
Finally, we define the equivalent constraint to the backup part (Eq.~\ref{eq:bDisj}).

\subsection{Capacity Constraints}


{\setlength{\parindent}{0cm}
\par{ {\it Node Capacity Constraints:}}
Virtual nodes from different tenants, mapped as a response to a different VNR, can be mapped to the same substrate node (in contrast to a node from the same tenant).
Let's call $\mathbb{N}^V$ the set of all virtual nodes that, at a certain moment, are mapped onto the substrate, and $ i \uparrow p $ to indicate that virtual node \textit{i} is hosted on the substrate node \textit{p}.
Then, the residual capacity of a substrate node, $ R_{N} (p) $, is defined as the CPU capacity currently available in substrate node $ p \in N^{S} $.
}

\begin{center} $ R_{N} (p) = cpu^{S}(p) - \smashoperator[r]{\sum_{\forall i \uparrow p}}cpu^{V}(i),~i \in \mathbb{N}^V$ 
\end{center} 

For a substrate node, we need to ensure that we never allocate more than its residual capacity, when carrying out a new embedding.
This needs to take into consideration the resources consumed by both the working and backup nodes (Eq. \ref{eq:capnoderesi}).

\begin{align}	
\smashoperator{\sum_{i\in N^{V}}} wn_{i,p} ~~ cpu^{V}(i)~+~\smashoperator{\sum_{j\in \overline{N}^{V}}}bn_{j,p} ~~ cpu^{V}(j) \leqslant R_{N}(p),~\forall p\in N^{S} \label{eq:capnoderesi}
\end{align}
    
{\setlength{\parindent}{0cm}
\par{ {\it Link Capacity Constraints:}}
Similarly, substrate links can also map virtual edges from different tenants. 
Lets define $\mathbb{E}^V$ as the set of all virtual edges mapped, at a certain instant, onto the substrate, and $ (i,j) \uparrow (p,q) $ to denote that the flow of virtual link $(i,j)$ traverses substrate link $(p,q)$.
The residual capacity of a substrate link, $ R_{E} (p,q) $, is defined as the total amount of bandwidth available on the substrate link $ (p,q)\in E^{S} $. 
}

\begin{center} 
$ R_{E} (p,q) = bw^{S}(p,q) - \smashoperator[r]{\sum_{\forall (i,j) \uparrow (p,q)}}bw^{V}(i,j),~(i,j) \in \mathbb{E}^V	$
\end{center} 

The following constraint ensures that the allocated capacity of a substrate link never exceeds the capacity of that physical link, taking into consideration both the working and backup parts.

\begin{align}	
	\smashoperator{\sum _{(i,j)\in E^{V}
}} wf_{p,q}^{i,j} + \smashoperator{\sum _{(i,j)\in \overline{E}^{V}}}bf_{p,q}^{i,j} \leqslant R_{E} (p,q),~\forall (p,q) \in E^{S} \label{eq:caplinkresi}
\end{align}

%% file: 04_evaluation.tex
\section{Evaluation} \label{sec:evaluation}

In this section we present performance results of our solution, considering diverse (virtual and substrate) network topologies, and diverse VNR settings.
Our evaluation aims to ask three main questions.
First, what is the performance of our solution when compared with the most common alternatives?
Second, how does the richer set of services offered by our solution (namely, security and availability) affect performance?
Third, can a multi-cloud virtualization provider benefit from offering these value-added services to its users?


\subsection{Experimental Setup} \label{subsec:simulationSetup}


The setting of our experiments follows the related literature on this problem.
Specifically, we have extended the simulator presented in~\citep{vineSimulator} to simulate the dynamic arrival of Virtual Network Requests (VNRs) to the system.
To create the substrate networks we resorted to the network topology generator GT-ITM~\citep{zegura1996gtitm}.
Two kinds of networks were evaluated: one based on random topologies, where every pair of nodes is randomly connected with a probability between 25\% and 30\%; and the other employing the Waxman model to link the nodes with a probability of 50\%~\citep{Naldi2005}. 

The substrate networks have a total of 25 nodes.
The CPU and bandwidth ($cpu^S$ and $bw^S$) of nodes and links is uniformly distributed between 50 and 100.
Each of these resources is associated with one of three levels of security ($sec^S \in \{1.0, 1.2, 5.0\}$), according to a uniform distribution.
The rationale for the choice of these specific values is to achieve a good balance between the diversity of security levels and their monetary cost.
For this purpose we performed an analysis of the pricing schemes of Amazon EC2 and Microsoft Azure, considering both ``vanilla'' and secure VM instances.
It was possible to observe a wide range of values depending on the security services included.
For example, while an EC2 instance that has container protection is around 20\% more expensive than a normal instance (hence our choice of $1.2$ for the intermediate level of security), the cost of instances that offer threat prevention or encryption is at least 5 times greater (our choice for the highest level of security).
The substrate nodes are also uniformly distributed among three clouds, each one with a different security level ($cloud^S \in \{1.0, 1.2, 5.0\}$) -- justified along the same line of reasoning.
The goal is to represent a setup that includes a public cloud (lowest level), a trusted public cloud, and a private datacenter (assumed to offer the highest security).


\newcolumntype{C}[1]{>{\centering\let\newline\\\arraybackslash\hspace{2pt}}m{#1}}
\begin{table}
\centering
\footnotesize
\def\arraystretch{1.5}
\setlength\tabcolsep{3pt}
\begin{tabular}{ |c| p{415pt}| }
  		\hline
 		\textbf{Notation} & ~~~~~~~~~~~~~~~~~~~~~~~\textbf{Algorithm description} \\
        \hline
  		\multirow{1}{*}{NS+NA} & SecVNE with no security or availability requirements for VNs\\
        \hline
  		\multirow{2}{*}{10S+NA} & SecVNE with VNRs having 10\% of their resources (nodes and links) with security requirements (excluding availability)\\
        \hline
        \multirow{1}{*}{20S+NA} & Similar to \textit{10S+NA}, but with security requirements (excluding availability) for 20\% of the resources\\
        \hline
        \multirow{1}{*}{NS+10A} & SecVNE with no security requirements, but with 10\% of the nodes requesting replication for increased availability\\
        \hline
        \multirow{1}{*}{NS+20A} & Similar to \textit{10S+NA}, but with availability requirements for 20\% of the resources\\
        \hline
        \multirow{1}{*}{20S+20A} & SecVNE with 20\% of the resources (nodes and links) with requirements for security and 20\% for replication\\
		\hline
        \multirow{1}{*}{D-ViNE} & VNE MILP solution presented in \citep{ViNEYard2012Boutaba} \\
		\hline
\end{tabular}
\caption{VNR configurations evaluated in the experiments.}
\label{tab:compAlgs}
\vspace{-0.2cm}
\end{table}

VNRs have a number of virtual nodes uniformly distributed between 2 and 4\footnote{Please note that a node corresponds to a virtual switch, that can support hundreds to thousands of containers in a large VM (recall Figure~\ref{fig:substrate}).
In fact, the setup of experiments with our prototype~\cite{Alaluna17} included running over 6 thousand containers per VM.}.
Pairs of virtual nodes are connected with a Waxman topology with probability 50\%.
The CPU and bandwidth of the virtual nodes and links are uniformly distributed between 10 and 20.
Several alternative security and availability requirements are evaluated, as shown in Table~\ref{tab:compAlgs}. 
We assume that VNR arrivals are modeled as a Poisson process with an average rate of 4 VNRs per 100-time units.
Each VNR has an exponentially distributed lifetime with an average of 1000-time units.

The MILP is solved using the open source library GLPK \citep{makhorin2008glpk}.
In the objective function, we set $ \beta_{1} = \beta_{2} = \beta_{3} = 1$ to balance evenly the cost components (Eq. \ref{eq:obj}).
Parameter $\alpha$ was also set to 1 because our pricing analysis showed negligible cost differences between intra- and inter-cloud links, in most of the relevant scenarios.
We set up 20 experiments, each with a different substrate topology (10 random and 10 Waxman).
Every experiment ran for 50~000 time units, during which embedding is attempted for groups of VNRs (specifically, 10 sets of 2~000 VNRs each were tested).
The order of arrival and the capacity requirements of each VNR are kept the same in each of the configurations of Table~\ref{tab:compAlgs}, ensuring that they solve equivalent problems.




In the evaluation, we compared our approach with the D-ViNE MILP solution~\citep{ViNEYard2012Boutaba}.
D-ViNE was chosen because it has been considered as the baseline for most VNE work~\cite{surveyEmbedding}, and due to the availability of its implementation as open-source software.
While D-ViNE requirements are only based on CPU and bandwidth capacities, our algorithm adds to these requirements security demands, namely node and link security (including availability), and cloud preferences. 

\subsection{Metrics} \label{sec:metrics}

We considered the following performance metrics in the evaluation:

\noindent \textit{-- VNR acceptance ratio:} the percentage of accepted requests (i.e., the ratio of the number of accepted VNRs to the total number of VNRs);

\noindent \textit{-- Node stress ratio:} average load on the substrate nodes (i.e., average CPU consumption over all nodes);


\noindent \textit{-- Link stress ratio:} average load on the substrate links (i.e., average bandwidth consumption over all  edges);


\noindent \textit{-- Average revenue by accepting VNRs:} One of the main goals of VNE is to maximize the profit of the virtualization provider. For this purpose, and similar to \citep{ViNEYard2012Boutaba, yu2008rethinking}, the revenue generated by accepting a VNR is proportional to the value of the acquired resources. As such, in our case, we take into consideration that stronger security defenses will be charged at a higher (monetary) value. Therefore, the revenue associated with a VNR is:  

\begin{center} 
$\mathbb{R}($VNR$)=\lambda_{1} \smashoperator{\sum_{i \in N^{V}}} [1 + \varphi_1(i)]~cpu^{V}(i)~~sec^{V}(i)~~cloud^{V}(i)$~+\\
$\lambda_{2} \smashoperator{\sum_{(i,j) \in E^{V}}} [1 + \varphi_2(i,j)]~bw^{V}(i,j)~~sec^{V}(i,j)$, 
\end{center}

\noindent where $\lambda_{1}$ and $\lambda_{2}$ are scaling coefficients that denote the relative proportion of each revenue component to the total revenue.
These parameters offer providers the flexibility to price different resources differently.
Variables $\varphi$ account for the need to have backups, either in the nodes ($\varphi_1(i)$) or in the edges ($\varphi_2(i,j)$).
Specifically, if $\varphi_1(i) = 1$ node $i$ requires backup, with $0$ otherwise; if $\varphi_1(i,j)=1$, at least one node $i$ or $j$ requires backup, with $0$ otherwise. 

This metric accounts for the average revenue obtained by embedding a VNR (i.e., the total revenue generated by accepting the VNRs divided by the number of accepted VNRs).


\noindent \textit{-- Average cost of accepting a VNR:}  The cost of embedding a VNR is proportional to the total sum of substrate resources allocated to that VN. In particular, this cost has to take into consideration that certain virtual edges may end up being embedded in more than one physical link (as in the substrate edge between nodes $b$, $d$ and $c$, in Figure~\ref{fig:exampleEmbedding}). The cost may also increase if the VNR requires higher security for its virtual nodes and links. Thus, we define the cost of embedding a VNR as:


\begin{center}
$\mathbb{C}($VNR$)=
\lambda_{1} \smashoperator{\sum_{i \in N^{V}}} ~\sum_{p \in N^{S}} cpu_{p}^{i}~~sec^{S}(p)~~cloud^{S}(p)
$~+\\$
\lambda_{2} \smashoperator{\sum_{(i,j) \in E^{V}}}~~\sum_{(p,q) \in E^{S}} f_{p,q}^{i,j}~~sec^{S}(p,q)$, \end{center}

\noindent where $ cpu_{p}^{i} $ corresponds to the total amount of CPU allocated on the substrate node $ p $ for virtual node $ i $ (either working or backup). Similarly, $ f_{p,q}^{i,j} $ denotes the total amount of bandwidth allocated on the substrate link $ (p,q) $  for virtual link $ (i,j) $. $\lambda_{1}$ and $\lambda_{2}$ are the same weights introduced in the revenue formula to denote the relative proportion of each cost component to the total cost.
In the experiments, we set $\lambda_{1} = \lambda_{2} = 1$.

\subsection{Evaluation Results} \label{subsec:evalRes}


Figure~\ref{evaluation:acceptanceRatioRandom} displays the acceptance ratio over time for one particular experiment with a random topology substrate. We can observe that after the first few thousand time units, the acceptance ratio tends to stabilize. A similar trend also occurs with the other experiments, and for this reason, the rest of the results are taken at the end of each simulation.   
Due to space constraints, the results in the following graphs are from the Waxman topologies only (Figure~\ref{evaluation:acceptanceRatio} - \ref{evaluation:revenue}).
We note however that the conclusions to be drawn are exactly the same as for the random topologies. 
The main conclusions are:

\begin{figure*}[t!]
  \centering
  \begin{subfigure}[b]{.32\textwidth}
    \centering
    \includegraphics[trim=.5cm 0 0 1cm, width=\linewidth]{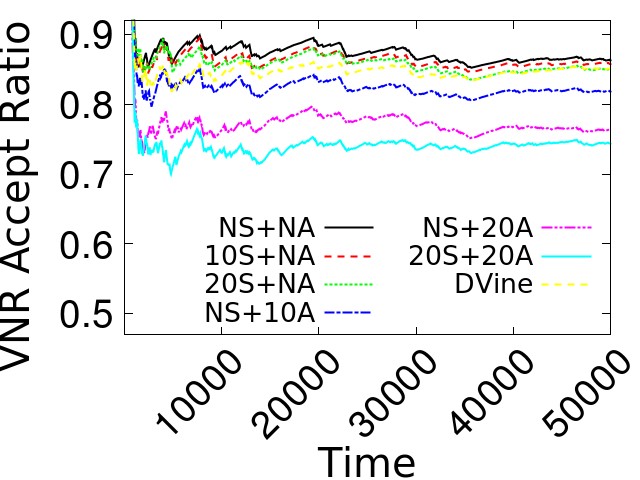}
    \caption{}
    \vspace{0.2\baselineskip}
    \label{evaluation:acceptanceRatioRandom}
  \end{subfigure}
   \begin{subfigure}[b]{.32\textwidth}
   \centering
   \includegraphics[trim=.5cm 0 0 1cm, width=\linewidth]{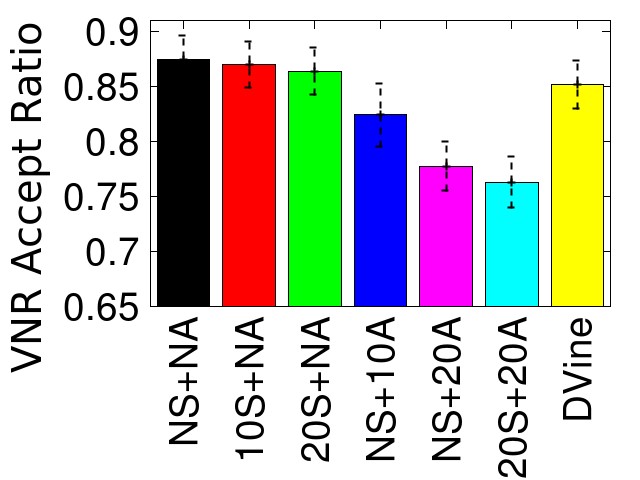}
   \caption{}
   \vspace{0.2\baselineskip}
   \label{evaluation:acceptanceRatio}
  \end{subfigure}
  \begin{subfigure}[b]{.32\textwidth}
   \centering
   \includegraphics[trim=.5cm 0 0 1cm, width=\linewidth]{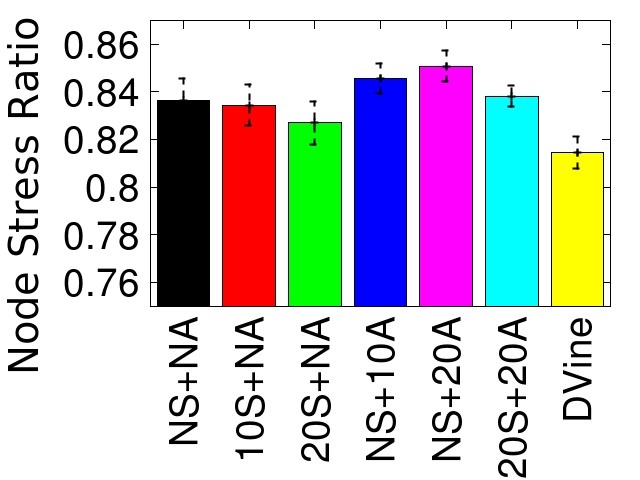}
   \caption{}
   \vspace{0.2\baselineskip}
  \label{evaluation:nodeStress}
  \end{subfigure}
  \centering
  \begin{subfigure}[b]{.32\textwidth}
   \centering
   \includegraphics[trim=.5cm 0 0 1cm, width=\linewidth]{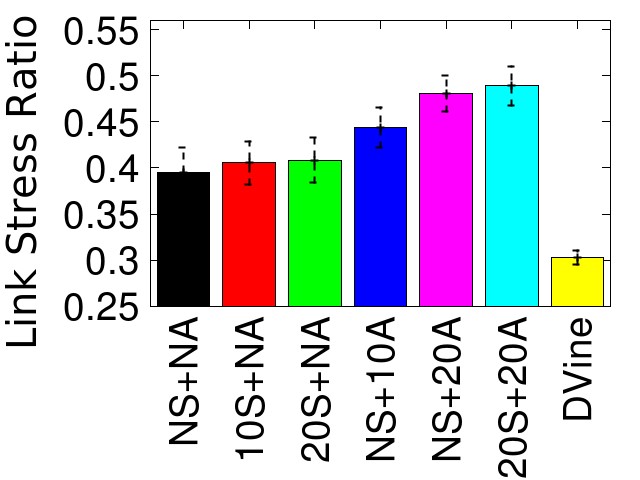}
   \caption{}
  \label{evaluation:linkstress}
  \end{subfigure}
  \begin{subfigure}[b]{.32\textwidth}
    \centering
    \includegraphics[trim=.5cm 0 0 1cm, width=\linewidth]{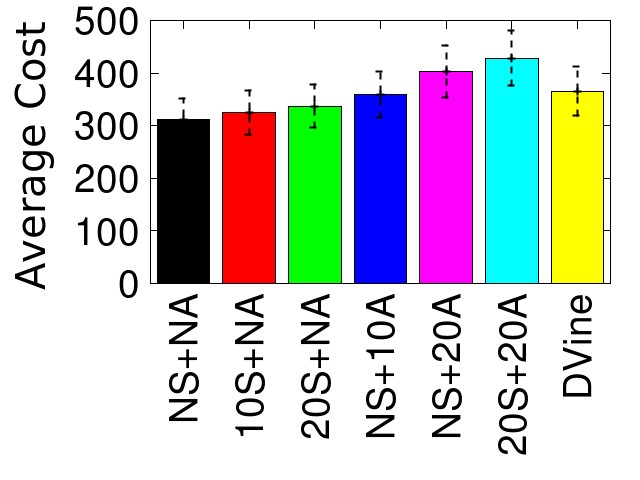}
    \caption{}
     \vspace{-0\baselineskip}
    \label{evaluation:cost}
  \end{subfigure}
   \begin{subfigure}[b]{.32\textwidth}
   \centering
   \includegraphics[trim=.5cm 0 0 1cm, width=\linewidth]{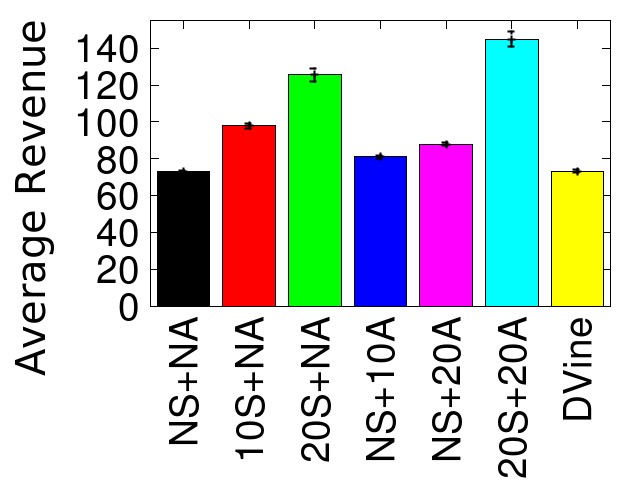}
   \caption{}
   \label{evaluation:revenue}
  \end{subfigure}
  \caption[]{Average results and standard deviation (except first graph): (a) VNR acceptance ratio over time; (b) VNR acceptance ratio; (c) Node utilization; (d) Link utilization; (e) Cost for accepting VNRs; (f) Revenue for accepting VNRs.}
   \label{fig:allMesures}
\end{figure*}

\vspace{0.2cm}
\noindent \textbf{SecVNE exhibits a higher average acceptance ratio when compared to D-ViNE, not only for the baseline case but also when including security requirements:} Figure~\ref{evaluation:acceptanceRatio} indicates that SecVNE can make better use of the available substrate resources to embed the arriving VNRs when compared to the most commonly employed VNE algorithm.
It is interesting to note that SecVNE is better than D-Vine even when 20\% of the VNRs include security requirements, which are harder to fulfill.
This does not mean D-Vine is a poor solution -- it merely shows that its model is not the best fit for our particular problem.
In particular, D-Vine uses the geographical distance of substrate nodes as one of the variables to consider in node assignment.
This parameter is less relevant in our virtualized environment but constrains D-Vine options.
In any case, notice that the results for D-Vine represent its best configuration with respect to geographical location -- we have tested D-Vine with the entire range of options for this parameter.

\vspace{0.2cm}
\noindent \textbf{A richer set of demands decreases the acceptance ratio, but only slightly:}
VNRs with stronger requirements have a greater number of constraints that need to be satisfied, and therefore it becomes more difficult to find the necessary substrate resources to embed them.
However, a surprising result is the small penalty in terms of acceptance ratio in the presence of security demands (see Figure~\ref{evaluation:acceptanceRatio} again).
For instance, an increase of 20 percentage points (pp) in the resources with security needs results in a penalty of only around 1 pp in the acceptance ratio.
Also interesting is the fact that the reduction in acceptance ratio is more pronounced when VNRs have availability requirements when compared to security.
In this case, an increase of 20 pp in the number of nodes with replication results in a penalty of around 10 pp. 
This is because of the higher use of substrate resources due to the reservation of backup nodes/links.

\vspace{0.2cm}
\textbf{Security demands only cause a small decrease on substrate resources utilization.}
Figures \ref{evaluation:nodeStress} and  \ref{evaluation:linkstress} show the substrate node and link stress ratio, respectively.
We observe that the utilization of node resources is very high in all cases (over 80\%), meaning the mapping to be effective.
It is also possible to see that slightly more resources are allocated in the substrate network with SecVNE than with D-ViNE, which justifies the higher acceptance ratio achieved.
If the existing resources are used more extensively to be able to serve more virtual network requests, then the assignment of virtual requests is being more effective.
As the link stress ratio is lower (again, for all cases), this means the bottleneck is the node CPU.
Finally, the link stress ratio of D-Vine is lower than in our solution.
This is due to D-vine incorporating load balancing into the formulation.

\vspace{0.2cm}
\textbf{Security and availability requirements increase costs and revenues.} 
Figures \ref{evaluation:cost} and  \ref{evaluation:revenue} display the average cost and revenue for each VNR embedding, respectively. The results show that reasonable increases in the security requirements (10\% and 20\%) only cause a slight impact on the costs. However, higher costs are incurred to fulfill availability needs due to the extra reservation of resources (nodes and links). Since D-vine does not consider security and availability aspects, it ends up choosing embeddings that are more expensive (e.g., with respect to ``NS+NA''). In terms of revenue, it can be observed that by charging higher prices for security services, virtualization providers can significantly enhance their income --- average revenue almost doubles for a 20\% increase in security needs.

%% file: 05_relatedWork.tex
\section{Related Work} \label{sec:relatedWork}


\textbf{Virtual Network Embedding.} There is already a wide literature on the VNE problem~\citep{surveyEmbedding}.
Yu et al.~\citep{yu2008rethinking} where the first to solve it efficiently, by assuming the capability of path splitting (multi-path) in the substrate network, which enables the computationally harder part of the problem to be solved as a multicommodity flow (MCF), for which efficient algorithms exist.
The authors solve the problem considering two independent phases -- an approach commonly used by most algorithms.
In the first phase, a greedy algorithm is used for virtual node embedding.
Then, to map the virtual links, either efficient MCF solutions or k-shortest path algorithms can be used.
In~\citep{ViNEYard2012Boutaba}, Chowdhury et al. proposed two algorithms for VNE that introduce coordination between the node and link mapping phases.
The main technique proposed in this work is to augment the substrate graph with meta-nodes and meta-links that allow the two phases to be well correlated, achieving more efficient solutions.
Neither of these initial works considered security.

\textbf{Availability in VNE.}
As failures in networks are inevitable, the issue of failure recovery and survivability in VNE has gained attention recently.
H. Yu et al.~\citep{yu2011costEfficient} have focused on the failure recovery of nodes.
They proposed to extend the basic VNE mapping with the inclusion of redundant nodes.
Rahman et al. \citep{rahman2010} formulated the survivable virtual network embedding (SVNE) problem to incorporate single substrate link failures.
Shahriar et al. \cite{Covine2016} went further to consider the presence of multiple substrate link failures.
Contrary to our work, these proposals target only availability.

\textbf{Multi-infrastructure VNE.}
The majority of the works in VNE field only consider a single infrastructure provider.
The exception is the work by Chowdhury et al.~\cite{Chowdhury2010polyvine}.
The problem which the authors address is, however, different from ours. 
They proposed PolyVINE, a policy-based end-to-end VNE framework, that considers the conflicts of interest between service providers, interested in satisfying their demands while minimizing their expenditure, and infrastructure providers, that strive to optimize resource allocation by accepting requests that increase their revenues while offloading less profitable work onto their competitors.
This work is thus orthogonal to the one we present here.

\textbf{Security in VNE.}
Another relatively less well-explored perspective on the VNE problem is providing security guarantees.
Fischer et al.~\cite{fischer2011secureVNE} have introduced this problem, considering the inclusion of constraints in the VNE formulation that take security into consideration.
The authors proposed the assignment of security levels to every physical resource, allowing virtual network requests to include security demands.
From this paper, we have borrowed the concept of security level as an abstraction to represent the different protection mechanisms from the substrate and the users' security demands.
This position paper has only enunciated the problem at a high level and has not proposed any solution or algorithm.
Bays et al.~\cite{Bays2012} and Liu et al.~\citep{liu2014secawarevne, Liu2015} have afterwards proposed VNE algorithms based on this idea.
None of these solutions support the detailed specification of security we propose.
The authors of~\cite{Bays2012} consider only link protection, by considering specific physical paths to provide a cryptographic channel, assuring confidentiality in communications.
Liu et al. go further, by considering also node security.
However, they also do not consider availability nor a multi-cloud setting with different trust domains.
In addition, their model makes a few assumptions that make it unfeasible for a practical network virtualization platform.
First, it requires virtual nodes to fulfill a specific security level, demanded by the physical host.
In practice, a virtualization platform cannot assume such level of trust about its guests. 
Second, it assumes that the duration of a virtual network request is known beforehand.
This limits its applicability in a traditional pay-as-you-go model, as a cloud tenant typically does not know in advance the duration of its requests.   
We make none of these assumptions in our work.
Finally, none of these works proposes a policy language to improve the expressiveness of substrate and virtual network specification.


%% file: 06_conclusion.tex
\section{Conclusions} \label{sec:conc}


In this paper, we presented a new solution for the secure virtual network embedding (SecVNE) problem.
Our approach fits a multi-cloud network virtualization deployment and enhances over the state-of-the-art on security.
Specifically, by allowing users to define different security and availability demands to the nodes and links of their virtual networks.
The solution further allows tenants to leverage a substrate composed of several clouds with distinct levels of trust, enabling specific node placement requirements to be set.
By not relying on a single cloud provider we avoid internet-scale single points of failures (with the support of backups in different clouds).
Besides, privacy and other security concerns can be accommodated by constraining the mapping of certain virtual nodes to specific classes of clouds (e.g., sensitive workloads can be placed in a private cloud).

We formulate the SecVNE model and solve it optimally as a Mixed Integer Linear Program (MILP).
In addition, we propose a new policy language to specify the characteristics of the substrate and virtual networks, improving the expressiveness of user requests.
Our extensive evaluation shows the efficiency of the solution and the favorable cost-revenue trade-offs of including security services in virtual networks.

\vspace{0.2cm}
\noindent {\bf Acknowledgments:} This work was partially supported by the EC through project SUPERCLOUD (H2020-643964), and by national funds of FCT with ref. UID/CEC/00408/2013 (LaSIGE) and ref. PTDC/CCI-INF/30340/2017 (uPVN project). Jos\'e Rui Figueira acknowledges the support from the FCT grant SFRH/BSAB/139892/2018.